# X-ray fluorescence spectra of metals excited below threshold

M. Magnuson, J.-E. Rubensson, A. Föhlisch[*], N. Wassdahl[**], A. Nilsson[***], and N. Mårtensson

*Department of Physics, Uppsala University, P. O. Box 530, S-751 21 Uppsala, Sweden*

**Abstract**

X-ray scattering spectra of Cu and Ni metals have been measured using monochromatic synchrotron radiation tuned from far above to more than 10 eV below threshold. Energy conservation in the scattering process is found to be sufficient to explain the modulation of the spectral shape, neglecting momentum conservation and channel interference. At excitation energies close to and above threshold, the emission spectra map the occupied local partial density of states. For the sub-threshold excitations, the high-energy flank of the inelastic scattering exhibits a Raman-type linear dispersion, and an asymmetric low energy tail develops. For excitation far below threshold the emission spectra are proportional to a convolution of the occupied and unoccuppied local partial densities of states.

## 1 Introduction

X-ray emission spectroscopy has a long tradition in solid state physics. The first experimental result that motivated the term Resonant Inelastic X-ray Scattering (RIXS) was the observation of x-ray emission of metals excited far below threshold by a conventional x-ray tube [1]. With the construction of tunable synchrotron radiation sources in the last decades, RIXS, also referred to as resonant x-ray Raman scattering, has become one of the most powerful techniques to study the electronic structure in solids as well as in atoms and molecules [2]. The x-ray scattering is a second-order optical process where the intermediate states are the same as the final states of x-ray absorption. Even though the intensity of the signal of a second-order optical process is generally much weaker than in a first order process such as photoemission, the obtained information is in many aspects useful. By tuning the incident photon energy, selected information about specific intermediate states can be obtained. This has led to new insights concerning assignments of features of different symmetries as well as the dynamics of the excitation-emission process. Experimentally, scattering into a localized intermediate state is manifested as a linear dispersion of the spectral features in the final states for varying excitation energies below threshold. The linear dispersion of the final states is often referred to as the Raman dispersion law [3]. Although it was early predicted that energy conservation in the scattering process should have observable consequences, this has been experimentally confirmed only in the last decades [2].
The characteristic feaures in RIXS spectra largely depend on the specific material properties. In materials with weak electron correlations such as in many semiconductors and insulators, experimental data are often well interpreted using electronic states based on density functional theory. For these kinds of broad band materials, a RIXS theory including resonance phenomena [4] and crystal momentum [5, 6] has been established. The aspect of the one-step scattering treatment is that the conservation of momentum is important for the overall process. This leads to a restriction of the excitations in the absorption-emission scattering process with strongly energy-dependent spectral structures observed ~ 0-25 eV above threshold and the prospects of partially performing band structure mapping is still under investigation [7].

When there are more than one excitation-emission path to the same final state the one-step nature of the process is manifested in interference effects. The scattering probability is then proportional to the square of the sum of the amplitudes corresponding to the various excitation channels. This is formulated more precisely by the Kramers-Heisenberg equation for inelastic x-ray scattering [8, 9]. For the radiative decay channel only the term corresponding to resonant anomalous scattering needs to be considered while the direct term and the non-resonant anomalous scattering term are neglected. If interference effects are important, information about the dynamics in the process can be extracted[10]. However, if only one single excitation-emission path dominates, interference effects can be neglected. Then a two-step picture is retrieved in the sense that the probability for scattering to a final state can be calculated by multiplying the probability for excitation to the intermediate state with the probability that this state decays to the final state.

In this paper, we investigate the spectral modifications of continuum excitations below the $2p$ core-level threshold in Cu and Ni metals. By applying a numerical model neglecting interference effects, we demonstrate that energy conservation is sufficient to explain the spectral modifications when the excitation energy is tuned far below the ionization limit. Similar observations both in the radiative [11, 12] and non-radiative [13, 14] decay channels in other systems have been presented as dynamic effects referring to the full scattering theory. It is emphasized that the selectivity in terms of energy of the incoming photons accounts for the spectral behaviour below threshold. The linear dispersion effect can be well described using a two-step approach if energy conservation is preserved. The model applied here is a two-step model in the sense that it neglects all interference and momentum conservation effects. However, is does apply energy conservation to the whole excitation-emission process. The results show that the relative probability for exciting an electron to higher energies is enhanced when the excitation energy decreases. We quantify this somewhat counter-intuitive result, and show that the features of the absorption spectrum are retrieved in the low-energy tail of the scattering spectrum. Far below threshold, the shape of the energy loss spectrum is well described as a convolution of the occupied and unoccupied local partial density of states.

## 2 Experimental Details

The measurements were performed at beamline 8.0 at the Advanced Light Source, Berkeley. X-ray absorption spectra recorded at the $2p$ absorption edges of the Cu(100) and Ni(100) single crystals were obtained in total electron yield by measuring the sample drain current. The Cu and Ni $L_{2,3}$ RIXS spectra were measured using a high-resolution grazing-incidence grating spectrometer with a two-dimensional position-sensitive detector [15, 16]. During the absorption measurements at the Cu and Ni $2p$ edges, the resolutions of the beamline monochromator were about 0.2 and 0.3 eV, respectively. RIXS spectra of the Cu and Ni crystals were recorded with resolutions better than 0.8 and 1.0 eV, respectively.

All the measurements were made at room temperature, with a base pressure better than $2 \times 10^{-10}$ Torr. In order to minimize self-absorption effects [17], the angle of incidence was about $\sim 7^o$ during the emission measurements. The emitted photons were always recorded in an angle perpendicular to the direction of the incident photons, with the polarization vector parallel to the horizontal scattering plane.

The cross section for exciting a core hole state around $\sim 10$ eV below a $2p$ threshold in transition metals is around 2000 times weaker than above threshold, suggesting that it would be impossible to obtain sufficient intensity for measurements. This is often the case in the non-radiative decay channel. However, the total fluorescence signal is not directly proportional to the scattering cross section. Since the probability that the photon is absorbed in the sample is close to unity, the variation in the cross section rather determines the depth distribution of fluorescence generation.

# 3 Results and Discussion

Figure 1 shows $L_3$ emission spectra (dots) of Cu metal excited at various photon energies from 932.5 eV at the Cu $L_3$ threshold down to energies as low as 921.2 eV. A spectrum excited at 1110 eV is also shown at the top. For a quantitative comparison of the spectral shapes at sub-threshold excitation, all the emission spectra were normalised to the same peak heights. Figure 2 shows corresponding $L_3$ emission spectra (dots) of Ni metal excited with photon energies from 852.7 eV at the Ni $L_3$ threshold down to energies as low as 840.7 eV (12.0 eV below threshold). A spectrum excited at 1096 eV is shown at the top and all the spectra were normalised to the same peak heights.

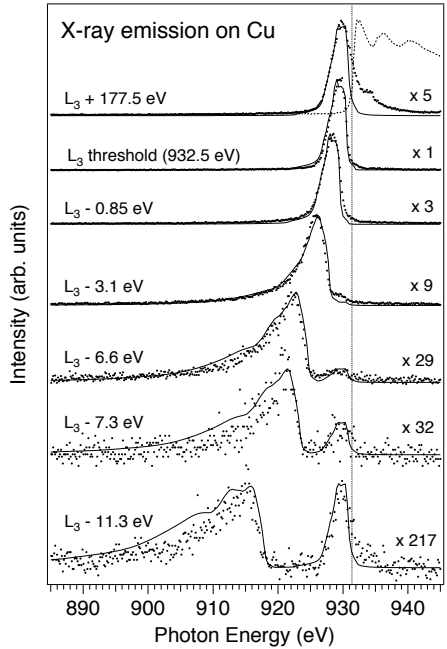

**Figure 1:** A series of emission spectra of Cu metal recorded at different excitation energies below and above the $L_3$ threshold. The dots and full curves are experimental and calculated RIXS data, respectively. The error bars are proportional to the noise of each spectrum. The dashed curve at the top is an XAS spectrum.

Although their spectral shapes are different, the Cu and Ni metal spectra basically show similar excitation energy dependences. Above threshold, the main $L_3$ line stays at constant emission energy. In the spectra excited far above threshold at 1096 eV and 1110 eV, for Cu and Ni, respectively, additional satellite structures are visible on the high energy flank of the main lines. For Cu, the energy-dependence of the satellite intensities above threshold have been discussed in terms of shake-up/shake-off and Coster-Kronig transitions in previous publications[18, 19].

In the following, we will describe the excitation-energy dependence of the emission spectra as the energy is tuned below the $L_3$ threshold in terms of a two-step model. It will be shown that most of the spectral modifications can be understood from simple energy arguments.

Energy conservation connects all states in the scattering process as

$$E_{in}+E_g=E_c+E_{e(c)}=E_h+E_{e(h)}+E_{out}. \qquad (1)$$

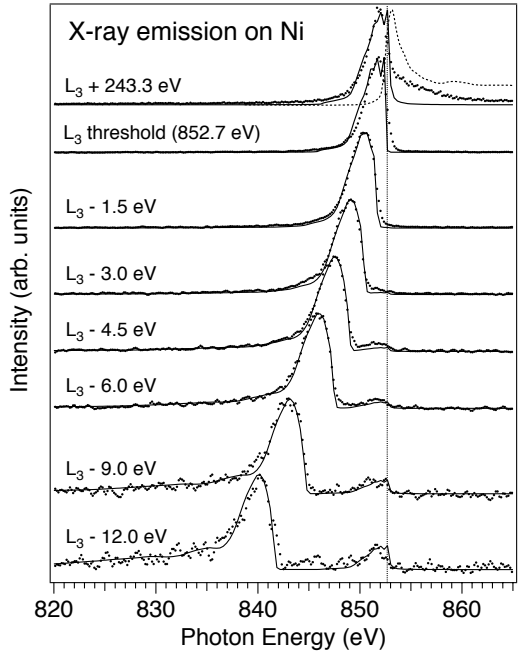

**Figure 2:** A set of emission spectra of Ni metal recorded at different excitation energies below and above the $L_3$ threshold. The dots and full curves are experimental and calculated RIXS data, respectively. The error bars are proportional to the noise of each spectrum. The dashed curve at the top is an XAS spectrum.

where $E_{in}$ is the energy of the incoming photon and $E_g$ is the energy of the ground state of the system. The sum of these energies is conserved in the scattering process, leading to core-hole creation with the core-hole state energy $E_c$ and the excited electron in the presence of the core-hole with the energy $E_{e(c)}$. Obviously, the total energy can be divided in different ways between $E_c$ and $E_{e(c)}$, depending in detail on the electronic structure of the studied material and the photon energy of the incoming photon $E_{in}$, leading to the occurrence of different intermediate states. As we are monitoring x-ray fluorescence, the final state we are interested in is then given by the energy sum of the outgoing photon $E_{out}$, the energy of the valence hole final state $E_h$ and the excited electron $E_{e(h)}$, now in the presence of the valence hole. We see directly that energy conservation allows for a given initial photon energy $E_{in}$ many scattering paths to reach a multitude of outgoing photon energies $E_{out}$, connecting the occupied and unoccupied electronic structure and leading to channel interference through multiple core-hole intermediate states reaching the equivalent final states. Furthermore, the core-hole intermediate states and the valence hole final states have through the Heisenberg uncertainty principle a natural lifetime energy $\Gamma$ due to their finite lifetime $\tau$, as $\Gamma\tau \geq \hbar$.

For continuum excitation, we can assume that the excited electron couples identically to the core-hole in the intermediate state as to the valence hole in the final state, then $E_{e(c)}=E_{e(h)}=E_e$, which is a significant simplification, schematically shown in Fig. 3. Then also interference effects can be neglected since only one excitation-emission path is possible to each final state. The two-step approximation is justified for determining the energy of the continuum electron. For excitation energies far below threshold, the lorenzian tails of the core-hole states lead to a non-selective scattering intermediate state involving all unoccupied states, which are dominated by the large number of continuum states.

The effect of the excitation of the wings of the Lorenzian functions gives rise to the highly asymmetric line profiles in the fluorescence spectra. It may be realized from Fig. 3 that the relative probability for exciting intermediate states with large $E_e$ becomes higher as $E_{in}$ decreases below threshold. Far below threshold, the selectivity in the excitation of various $E_e$ is lost. Energy conservation gives the energy of the outgoing photon:

$$E_{out} = E_{in} - (E_h + E_e - E_g). \tag{2}$$

The energy of the intermediate state does not enter into the equation. In principle, the widths of the spectral features are not limited by the natural width of the intermediate state and $E_{out}$ varies linearly with $E_{in}$. For a metal this is not directly observed because the final state energy is shared between $E_h$ and $E_e$, so that several final states are involved for each definite energy loss, $E_{in} - E_{out}$.

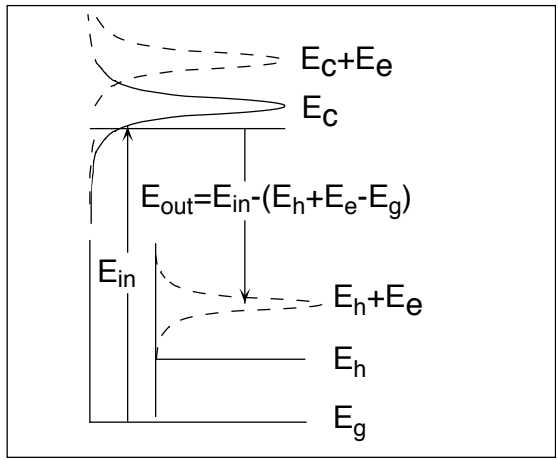

**Figure 3:** Schematic illustration of continuum excitation of Lorenzian tails below a core-level threshold.

However, there are certain restrictions on $E_h$ and $E_e$. The kinetic energy of the electron is non-negative, and with the energies defined in Fig. 3, $E_h - E_g \geq 0$. Eq. 2 then puts an upper limit to $E_e$ so that $0 \leq E_e \leq E_{in} - E_{out}$. If the excited electron does not couple to the rest of the system we can use a two-step approximation for the excitation-emission process. We assume that the energy of the excited electron, $E_e$, is the same after the second (emission) step as after the first (excitation) step so that

$$E_{in} = E_e + (E_c - E_g), \tag{3}$$

where $E_c$ is the energy of the core hole state. If $E_{in}$ is well-defined, Eq. 3 implies that $\partial E_e = -\partial E_c$, where $\partial$ is the partial derivative, i.e., it depends on a specific variable. Consequently, a lifetime broadening of the core hole state leads to an uncertainty in the energy of the excited electron. In the case of a well-defined $E_h$ we see from Eq. 2 that $\partial E_{out} = \partial E_e$, and consequently $\partial E_{out} = -\partial E_c$. Thus, for a metal the lifetime broadening of the intermediate state does limit the width of the spectral features via the uncertainty in $E_e$ in the final state.

If $E_c$ is independent of $E_{in}$, it is observed in Eq. 3 that a variation in $E_{in}$ results in a variation of the excess energy which goes into the excited electron $E_e$. Combining Eqs. 2 and 3 we have

$$E_{out} = E_c - E_h, \tag{4}$$

which is the energy expression for non-resonant x-ray emission. Thus, $E_{out}$ seems to be independent of $E_{in}$. This independence (Eq. 4) is strict, however, only if $E_c$ is well-defined. Due to the finite lifetime, the intermediate core hole state is better represented by a Lorentzian energy distribution centered at $E_c$ with a width proportional to the total decay rate. A strict upper limit to $E_{out}$ is given by Eq. 2 since $E_e \geq 0$. Thus, the high-energy cut-off in the spectrum ($E_{out}^{max}$) does show a linear dispersion with $E_{in}$ according to the Raman law. This is hardly noticeable in spectra excited above threshold, because only an insignificant part of the upper tail of the energy distribution is cut. However, in a recent study of the non-radiative channel in Cu, these effects have been shown[14]. Below threshold ($E_{in} < (E_c - E_g)$), on the other hand, only the lower tail of the energy distributions can give a contribution to the spectrum. Therefore, the upper limit given by Eq. 2 becomes crucial. This implies that the spectrum develops an asymmetric low-energy tail with decreasing $E_{in}$. Assuming a well-defined $E_h$, it is directly observed in Eq. 2 that the intensity at $E_{out} < E_{out}^{max}$ corresponds to $E_e > 0$.

The cross section for exciting a core electron to an excited state with energy $E_e$ by a photon beam of energy $E_{in}$ is given by

$$P_{exc}(E_{in}, E_e) \propto \frac{\varrho_u(E_e)}{((E_{in} - (E_c + E_e - E_g))^2 + (\frac{\Gamma_c}{2})^2)} \tag{5}$$

where $\varrho_u(E_e)$ is the weighted unoccupied partial DOS and $\Gamma_c$ is the total decay rate of the intermediate core hole state. The probability that this core excited state decays to a final state with a valence hole of energy $E_h$ and an electron of energy $E_e$ is simply

$$P_{emi}(E_{in}, E_{out}, E_e) \propto P_{exc}(E_{in}, E_e) \times \varrho_o(E_h - E_g), \tag{6}$$

where $\varrho_o(E_h)$ is the weighted occupied partial DOS. The prediction of the spectrum, $P_{emi}(E_{in}, E_{out})$, is now achieved by using Eq. 2 and integrating over all possible $E_e$:

$$P_{emi}(E_{in}, E_{out}) \propto \int_0^{E_{in} - E_{out}} \partial E_e \times G \times \tag{7}$$

$$\frac{\varrho_u(E_e) \times \varrho_o(E_{in} - (E_{out} + E_e))}{((E_{in} - (E_c + E_e - E_g))^2 + (\frac{\Gamma_c}{2})^2)}$$

where the $L_3$ core-hole lifetimes $\Gamma_c$ used in the simulations were 0.41 and 0.31 eV for Cu and Ni, respectively[20]. The occupied and unoccupied bands were modelled by the ground state partial 3$d$-DOS, $\varrho_o$ and $\varrho_u$[21]. The monochromator function G, was assumed to be a Gaussian centered at the nominal excitation energy plus a wide Lorenzian representing a high-energy leakage. About 0.1 % of the intensity at the nominal energy was needed in order to reproduce the features at constant energies below threshold at 930 and 852 eV for the Cu and Ni emission spectra, respectively. This is an effect of a non-perfect monochromator function and similar type of structures in other systems have been termed Stokes-Doubling [22]. The results of the simulations are presented in Fig. 1 and 2 as full curves. The overall agreement between the experimental data and the simulations is generally excellent. The highly asymmetric line profiles at large detuning energies are well reproduced.

The numerical model used here describes the spectra far above- as well as below- threshold excitation and there is no difference in the underlying physics. The excitation energy dependence is directly related to the selectivity of $E_e$ in the final state for a certain $E_{in}$. As observed in Eqn. (7) for every $E_{in}$ above threshold $E_{in} \geq (E_c - E_g)$ the Lorentzian function has a maximum for a certain $E_e$. For excitation below threshold, on the other hand, the Lorentz factor in Eq. (7) has no maximum for non-negative $E_e$. Instead the variations in $\varrho_u$ become important for the definiton of $E_e$. For excitation energies very far below threshold, the denominator in Eq. (7) varies little compared to $\varrho_u$ and consequently the spectral profile becomes a convolution between $\varrho_u$ and $\varrho_o$.

The analysis above does not include momentum conservation. This differs from previous analyses[11] of sub-threshold excited spectra which are based on momentum conservation for the overall scattering process[4, 6]:

$$\mathbf{Q} = \mathbf{k}_e - \mathbf{k}_h + \mathbf{G}, \qquad (8)$$

where $\mathbf{k}_e$ and $\mathbf{k}_h$ are the momenta of the electron and the hole, $\mathbf{Q}$ is the momentum transferred by the photons and $\mathbf{G}$ is a reciprocal lattice vector added in the reduced Brillouin zone. If $\mathbf{k}_e$ is well-defined, emission is allowed only at those $\mathbf{k}_h$ that fulfil Eq. 8 and the band-structure E($\mathbf{k}$) can be partially retrieved, in broad band materials. Thus, whereas $E_e$ is irrelevant for the emission $\mathbf{k}_e$ is not. For above threshold excitation a certain $\mathbf{k}_e$-selectivity thus gives an additional excitation energy dependence by restricting the possible final states.

At excitation energies below threshold the momentum selectivity is lost since all $\mathbf{k}$-values are excited simultaneously. Far below threshold, momentum selectivity can be completely neglected and the joint density-of-states (JDOS) is measured [11, 12]. However, the model applied here is equally important since the restrictions imposed by Eq. 8 in most cases are relaxed. This is due to transfer of momentum by additional electron excitations, phonons and other particles involved which are not measured in the experiment. Furthermore, at relatively high photon energies e.g., at the 2$p$ core levels of the late transition metals, there will be non-vertical transitions unless the incoming and outgoing photons are parallel [23]. This implies that the momentum transfer has to be considered when constructing the generalized JDOS. At lower photon energies, the momentum transfer can be neglected and the ordinary JDOS measured.

It has previously been argued that momentum conservation should be irrelevant in metal spectra, due to relaxation in the intermediate state. Indeed, electron and phonon coupling also set the limit for the application of Eq. 8. Yet, if multiparticle excitations are involved in the excitation step, we can replace $E_e$ with a general excess energy, and the one-electron DOS with a general DOS. Crucial for the derivation is only that the excess energy created in the first excitation step is unaffected by the second emission step (Eq. 3). Eq. 8 is much more sensitive to such effects, and the momentum of the electron cannot be generalized in the same way as $E_e$. In resonant scattering over discrete states a gradual change of the resonant spectra to a shape that simulates the non-resonant case as the excitation energy is detuned from the resonance has been observed [24, 25]. However, in the present case this is not equivalent to the nonresonant case since the selection rules are projecting the *s* and *d* character of the states. The core-hole angular-momentum symmetry puts a restriction on the final states. A strict comparison between the predictions of the generalized JDOS with a convolution of the DOS for systems with suitable absorption resonances at lower excitation energies where the momentum transfer is smaller would put this to a test.

# 4 Conclusions

Soft X-ray fluorescence spectra of the Cu and Ni transition metals have been measured using monochromatic synchrotron radiation. Due to the sensitivity of the fluorescence technique, $L_3$ sub-threshold spectra show a remarkably strong signal. The high-energy flank of the emission disperses linearly with the excitation energy and a pronounced low energy tail develops. This is interpreted as a consequence of the enhanced relative probability of exciting continuum electrons at higher energies when the excitation energy is lowered. Simulations of the spectra reveal that for large detuning energies, the spectra are largely modulated by the partial density of empty states and basically appear as a convolution of the occupied and unoccupied local partial density of states.

# 5 Acknowledgments

This work was supported by the Swedish Research Council, the Göran Gustafsson Foundation for Research in Natural Sciences and Medicine, the Swedish Institute (SI) and the Swedish Foundation for International Cooperation in Research and Higher Education (STINT). ALS is supported by the U.S. Department of Energy, under contract No. DE-AC03-76SF00098.